\documentclass[prb,twocolumn,showpacs,preprintnumbers,amsmath,amssymb]{revtex4}

\usepackage{graphicx}% Include figure files
\usepackage{dcolumn}% Align table columns on decimal point
\usepackage{bm}% bold math
\usepackage{epstopdf}
\usepackage{color}

\begin{document}

\title{Plastic response by dislocation glide in solid helium under dc strain rate loading}

\author{Caizhi Zhou$^{1,2,3}$}
\email{zhouc@mst.edu}
\author{Jung-Jung Su$^{4}$}
\author{Matthias J. Graf$^1$}
\author{Charles Reichhardt$^1$}
\author{Alexander V. Balatsky$^{1,5}$}
\author{Irene J. Beyerlein$^1$}
\affiliation{$^1$Theoretical Division, Los Alamos National Laboratory, Los Alamos, New Mexico 87545, USA \\
$^2$Center for Nonlinear Studies, Los Alamos National Laboratory, Los Alamos, New Mexico 87545, USA \\
$^3$Dept.\ of Materials Science and Engineering, Missouri University of Science and Technology, Rolla, MO 65409, USA\\
$^4$Dept.\ of Electrophysics, National Chiao Tung University, Hsinchu 300, Taiwan\\
$^5$Nordita, 23 Roslagstullbacken, Stockholm, Sweden}
\date{\today}

\begin{abstract}

We develop a model for the gliding of dislocations and plasticity in solid $^4$He.
This model takes into account the Peierls barrier, multiplication and interaction of dislocations, as well as
classical thermally and mechanically activated processes leading to dislocation glide.
We specifically examine the dc stress-strain curve and how it is affected by temperature, strain rate, and dislocation density.
As a function of temperature and shear strain, we observe plastic deformation and discuss 
how this may be related to the experimental observation of elastic anomalies in solid hcp  $^4$He 
that have been discussed in connection with the possibility of supersolidity or giant plasticity.
Our theory gives several predictions for the dc stress strain curves, for example, the yield point and the change in the work-hardening rate and plastic dissipation peak, 
that can be compared directly to constant strain rate experiments and thus provide bounds on model parameters.

\end{abstract}


%\pacs{67.25.dt, 67.80.B-, 67.80.bd}% PacS, the Physics and Astronomy
                             % Classification Scheme
% 67.80.bd --supersolid He-4
%67.80.-s -- quantum solids
%67.80.B- -- He-4 solid phase
%67.25.dt -- He-4 sound in
%61.72.Hh -- slip (dislocations)

\maketitle

\section{Introduction}
Studies in solid $^4$He  have recently received renewed interest due to the observation of 
a period drop in torsional oscillator experiments.\cite{Chan04,Rittner06,Kondo07,Aoki07,Penzev08,Hunt09,Pratt11}
This drop was interpreted as possible evidence for a supersolid phase.
Subsequently,
the same system was shown to exhibit an anomalous softening of the shear 
modulus with increasing temperature in the same temperature regime
where the period drop was observed.
It suggested that the shear modulus change and the period drop were    
connected and dislocations which contribute to the shear modulus softening played
an important role in the possible supersolid behavior.\cite{Day07, Day09, West2009}
One interpretation of how dislocation dynamics produce the period drop anomaly
is that  
the dislocations are not freely gliding but are pinned at certain points 
while the unpinned portions
vibrate, and that these vibrations produce additional damping.\cite{Day07, Day09, Mukharsky2009, Syshchenko10, Rojas11,West2009}
More recently, there have been more precise follow-up experiments on solid $^4$He in porous Vycor glass, 
where the period drop in the torsional
oscillator measurements was no longer observed, \cite{DYKim2012} 
arguing against a supersolid scenario; 
however, a still open question is what is responsible for the observed 
anomaly in the shear modulus. 

Although the issue of supersolidity may be close to being settled,
solid $^4$He still presents interesting questions in terms 
of the mechanical response
of materials with dislocations, and may exhibit elastic-plastic properties 
that are much more difficult to access in conventionally studied 
materials systems.  
We propose that the possible cause for the change of elastic properties measured in experiments 
is plastic deformation arising from the gliding motion of dislocations. 
This is in contrast to 
the vibrating string model of dislocations pinned by a network that 
do not show true (free) glide.\cite{Iwasa1979,Paalanen1981,Hiki1983,Iwasa2010,Iwasa2013}
The important questions are whether the anomaly
is caused by the free glide of dislocations, 
whether the model is consistent with the observed experiments, 
and whether the model can provide additional predictions. 
In our model, dislocation glide 
produces clear signatures in dc stress-strain curves in solid 
$^4$He similar to the effects established for other materials, 
such as dc stress-strain curves for
 metals that are discussed in the 
literature.\cite{Mecking1981,Orowan1954,Friedel1967,Nabarro1967,Hull2011}

The study of dislocation motion and plastic deformation of crystals has a long 
tradition and is 
typically analyzed by   
conducting dc
stress-strain curves for different strain rates and 
temperatures.\cite{Mecking1981,Essmann79,Beyerlein08,Brown2012} 
Here we develop a dislocation dynamics model based on rate-controlled 
mechanisms, 
where thermally-activated flow is assumed with basal or prismatic slip in hcp crystals (like solid $^4$He at low temperatures). 
In addition, the theory accounts for the creation and removal of dislocations as well as their glide. Since mobile dislocations get
pinned many times and at many points along their length due to the interactions with other dislocations, the effect of dislocation interaction must be included explicitly. Finally, the rate of plastic strain is controlled by the rate at which thermal and mechanical energy assist these segments to
overcome their energy barriers, allowing the rest of the dislocation line to bow out and spread before it is stopped again. 
\cite{Hirsch1959, Friedel1963, Saada1963, Seeger1963,Friedel1967,Nabarro1967,Kocks75} 

Many of the recent studies of the elastic properties of solid $^4$He were performed with ac measurement techniques of alternating shear strain, 
where the piezoelectric transducers were embedded in the solid helium. 
Here we focus instead on dc drives where we can obtain
clear predictions for the dislocation glide, 
since under a dc drive the dislocations
can move significant distances in the sample. 
Under an ac drive, the dislocations may not glide very far and 
it would be difficult to distinguish the
smaller ac motion from the vibration of a dislocation segment.

In the theory of plastic deformation of crystals, the role of dislocations gliding in specific crystallographic directions on specific crystallographic slip planes is related to a change of the work-hardening rate of the solid with temperature and strain rate. \cite{Taylor1934,Orowan40,Wielke76,Kocks75,Kocks03}
The hypothesis that this microscopic picture of slip and propagation of dislocations is relevant to solid helium received additional support from recent experiments in the hcp phase of solid $^4$He. \cite{Fefferman2012} 
More recently the reported low-temperature phenomena were interpreted as effects of giant plasticity caused by gliding dislocations. \cite{Haziot2013a,Haziot2013b}
Naturally, the same mechanism may also be responsible for similar anomalous behavior in disordered bcc crystals of $^4$He at temperatures close to the melting line,\cite{Eyal12a,Eyal12b} and in hcp $^3$He.\cite{West2009}

In this paper, we develop a dislocation glide model for solid $^4$He that is motivated by earlier work of Kocks and coworkers \cite{Kocks75,Mecking1981,Kocks03} 
on the dynamics of mobile dislocations moving through a dislocation {\it forest}.
This model is based on the thermodynamics and kinetics of dislocation glide to describe plastic deformation of crystals. 
However, we extend the framework of this classical plasticity model by interpreting the zero-point motion and quantum effects of solid helium as renormalization of potentials. This will not qualitatively change the kinetics and dynamics, but instead effectively renormalize certain thermodynamic quantities, such as by lowering the pinning potential, Peierls barrier, or screening of the dislocation-dislocation interaction.\cite{Kantha1994,Zaanen2004,Proville} 
In the case of the dc shear strain rate experiment, $\dot{\epsilon}=const.$, with shear stress $\tau$, we predict the softening of the work-hardening rate (WHR), $d\tau/d\epsilon$, with increasing temperature $T$ due to dislocation glide and the creation and multiplication of dislocation loops. Our dislocation glide model also predicts that higher $^3$He impurity concentration in solid $^4$He pushes the onset of the magnitude change in the WHR to higher temperatures by assuming that more $^3$He atoms pin more dislocations. At the same time, the zero-temperature value of the WHR is unaffected by immobile dislocations. Finally, our calculations show that at zero temperature the WHR equals the elastic shear modulus independent of applied strain, while at finite temperature it decreases with increasing strain to a saturated value.\cite{Zhou12}

We note that our theory is classical and the equations presented in section IIA are classical. 
It may be possible that quantum effects would come into play in our model, 
but they would only renormalize the parameters without altering the equations.
In this way, including quantum effects in our model is consistent with the 
classical theory that describes classical solids
as discussed in the wide body of literature on plasticity in classical systems.      

We account for thermal and quantum fluctuations in the thermodynamic part of the model with renormalized phenomenological model parameters of the energy barrier. 
A key result of this work is the prediction of a dissipation peak caused by plastic deformation of the solid in the same temperature region where the WHR changes most rapidly. 
The similarity of the dc strain rate model predictions to existing shear measurements of applied ac strain rate is remarkable, \cite{Day07, Day09, West2009, Syshchenko10, Rojas11} although a direct comparison is not possible due to significantly different loading and reverse-loading histories. Therefore, our results for the proposed dc strain rate experiments will provide stringent tests for the study of the plastic response of dislocation glide in the classical versus quantum regime at low temperatures of solid $^4$He.

\section{Plastic glide of dislocations}

\begin{figure}[t]
\rotatebox[origin=c]{0}{\includegraphics[width=0.70\columnwidth]{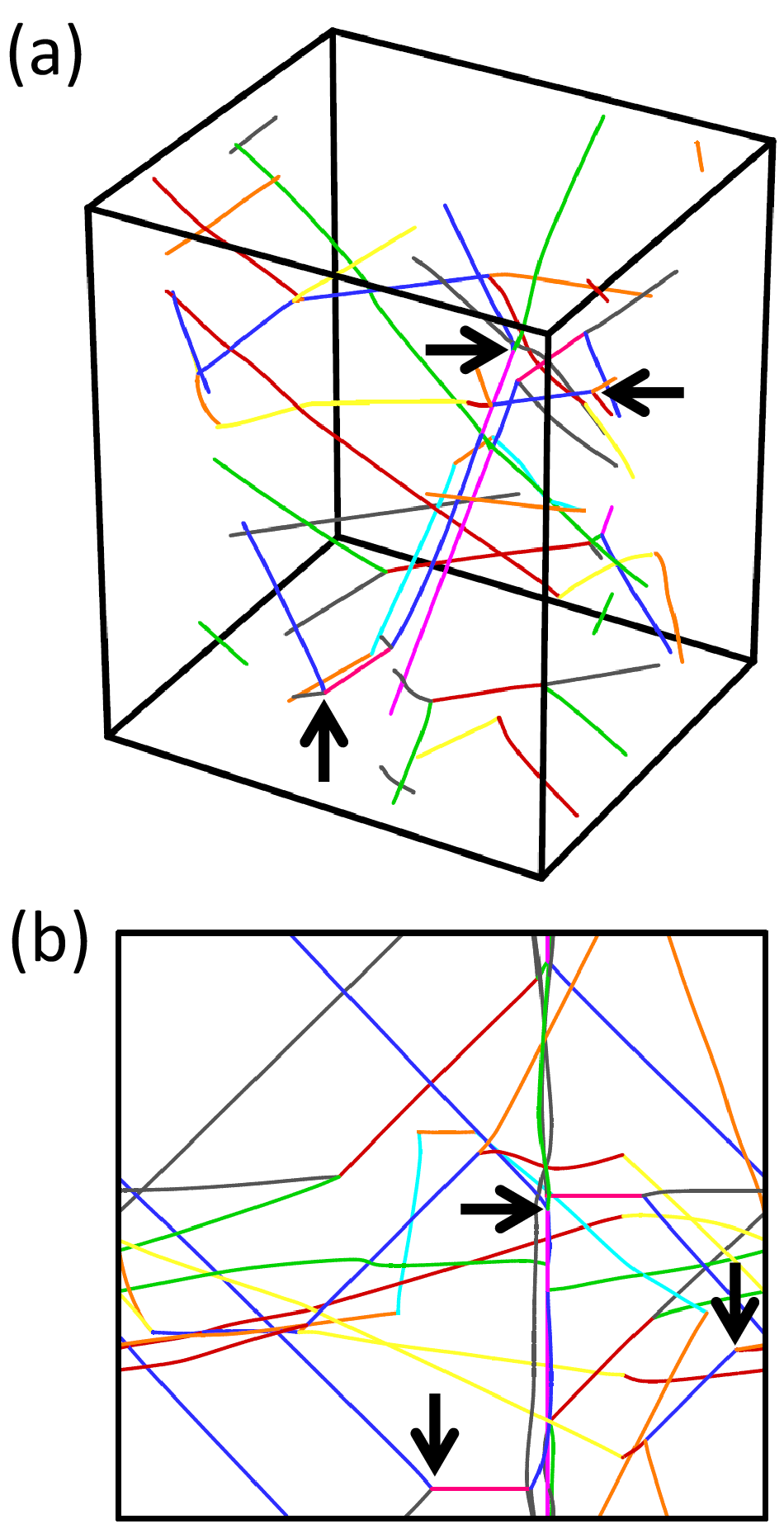}}
\caption{(color online). 
Characteristic dislocation network in crystalline materials  illustrates the complexity of dislocation glide.
(a) Snapshot of the simulation of typical three-dimensional (3D) network of dislocations in crystal. Different colors represent different Burgers vectors 
that indicate the magnitude and direction of strain imparted by glide of the dislocations.
(b) Top view of 3D dislocation structure. Pinning points marked by arrows will unzip from junctions under external loading;
see the SM for details on the simulation and the animation. \cite{SM}
}
\label{fig:FIG1}
\end{figure}

In crystalline materials like $^4$He, plastic deformation can arise from glide of linear structural defects called dislocations. Dislocations are commonly found in $^4$He crystal, even in the {\it cleanest} samples, \cite{Rojas11} due to the growth and cooling processes. A typical dislocation structure similar to that in metals is assumed in our model. In Fig.~\ref{fig:FIG1} we show a snapshot of the complexity of such a three-dimensional (3D) dislocation network commonly studied in metals, see the Supplemental Material (SM) \cite{SM} for details on the simulation and the animation of the dislocation glide, as well as the unzipping of pinning points marked by arrows. 
We borrow various elements from the seminal works on evolution laws of thermally activated dislocation glide
by Essmann and Mughrabi,\cite{Essmann79}  and Kocks and collaborators.\cite{Mecking1981,Kocks03}
and categorize the dislocations as being either {\it stored} or  {\it mobile}. The stored dislocations can assemble into networks inside individual crystals and do not contribute to the plastic deformation. Hence an increase in pinned or stored dislocations increases the WHR of the solid at higher temperatures.
The mobile dislocations, on the other hand, give rise to plastic strain and dissipation. However they may not stay in their own category. When a mobile dislocation glides through the network of stored dislocations it can be pinned and become stored. This pinning can nevertheless be overcome by thermal activation and applied strain and the dislocation then becomes mobile again.

\subsection{Dislocation glide model}

In our dislocation glide model, we adopt a statistical representation in which individual dislocation characteristics (mobility, polarity, screw/edge character, line orientation, and position) are not treated explicitly. Of course, this is a great simplification of the complex problem of dislocations and presents the first step toward capturing the essential mechanisms and physics underlying dislocation glide in solid helium, similar to studies of other materials in the past.
For simplicity, the large network of dislocations is represented as a continuous distribution of dislocation lines within a representative volume, characterized by a dislocation density with units of total dislocation line length per volume. The {\it total} dislocation density in a material at time $t$ consists of both {\it mobile} and {\it stored} (immobile) dislocations 
\begin{eqnarray}\label{eq:tot}
\rho_{tot}(t)=\rho_{mob}(t)+\rho_{stored}(t).
\end{eqnarray}

In deformed crystals, stored dislocations assemble into networks within the interior, as shown in Fig.~\ref{fig:FIG1}. Here, we assume that stored dislocations take on a similar three dimensional pattern in strained solid helium.  Accordingly, mobile dislocations moving in any direction and on any slip plane are likely to become immobile, when they encounter a {\it forest} of stored dislocations intersecting the glide plane, with which they strongly interact.
When dislocations glide, it was proposed that the total dislocation density increases by\cite{Taylor1934,Orowan40,Kocks03}
\begin{eqnarray}\label{eq:eom}
\Delta\rho_{tot}(t)=\Delta\epsilon_{p}(t)/{\it b}L ,
\end{eqnarray}
per unit time $\Delta t$. Here $b$ is the magnitude of the Burgers vector, $L$ is the mean-free path of mobile dislocations in a forest of dislocations, which is assumed to equal the interdislocation distance $1/\sqrt{\rho_{tot}}$, and $\Delta\epsilon_{p}$ is the increment of plastic strain produced by dislocation glide. 
The increments in stored and mobile dislocation densities are generally proportional to $\Delta\rho_{tot}$. As portions of an expanding dislocation loop can react with other loops and become immobilized it changes the density of stored dislocations by $\Delta \rho_{stored}$.
In this study, we assume for simplicity that newly added dislocations have equal chance of becoming mobile or stored. 
Thus, we arrive at  the change per unit time $\Delta t$: 
$\Delta\rho_{mob}=\Delta\rho_{stored}=\Delta\rho_{tot}/2.$

Mobile dislocations glide freely with a velocity distribution, which we parameterize only with its average speed $v$ for convenience.   In our dislocation network picture, see Fig.~\ref{fig:FIG1}, a dislocation can be pinned with potential  $U_i$ at local points, where it encounters other dislocations. We assume that the dislocation overcomes the pinning potential and glides by the combined effect of thermal activation (Arrhenius term) and external drive $\tau_{app}$. This assumption is supported by the experimental observation that mechanical stress increases with applied strain and decreases with temperature. \cite{Kocks03,Kocks75,Wielke76} As the dislocation glides, the whole line still undergoes resistance from the Peierls stress $\tau_{Peierls}$ and the mean-field dislocation-dislocation interaction stress $\tau_{int}$. The average speed is thus described by a combination of thermodynamic potential and attempt frequency for overcoming the barrier\cite{Kocks75,Kocks03}
\begin{eqnarray}\label{eq:velocity}
v(t) &=& b \nu_D \exp \left[ -\Delta G(t)/k_B T \right]  , 
\\
\Delta G(t) &=&  U_i - \big[ \tau_{app}(t)-\tau_{Peierls}-\tau_{int}(t) \big] V ,
\end{eqnarray}
where $\nu_D$ is the Debye or attempt frequency and $V=b^2  \Lambda$ is  the activation volume. $\Lambda$ is assumed to be the average distance between pinning points, which equals the interdislocation distance $L$, and $k_B$ is the Boltzmann constant.
The product $b \nu_D$ is a typical sound speed of the solid, and
$\Delta G$ is the thermodynamic Gibbs free energy for thermal activation of glide. We model the interaction between dislocations to be long ranged. In mean-field theory the interaction is inversely proportional to the interdislocation distance of a randomly oriented dislocation forest, which results in the resistive interaction stress\cite{Kocks75,Mecking1981,Lavrentev1980,Kocks03}
\begin{eqnarray}\label{eq:resist}
\tau_{int}(t)=\alpha \mu b \sqrt{\rho_{tot}(t)},
\end{eqnarray}
with elastic shear modulus $\mu$ and dislocation interaction coefficient $\alpha$.
Indeed, the dependence of $\tau_{int}$ on $\sqrt{\rho_{tot}}$ has been observed in many materials.
\cite{Staker1972,Kocks75,Kocks03}

We can now relate the macroscopic strain quantity with the above mentioned microscopic quantities of dislocation motion.
The total applied strain is composed  of both elastic and plastic strain, $\epsilon_{app} \equiv \epsilon_{e}+\epsilon_{p}$, while 
the plastic strain rate due to glide obeys the kinematic equation relating microscopic and macroscopic quantities\cite{Orowan40,Kocks03}
\begin{eqnarray}\label{eq:strain_rate}
\dot{\epsilon}_p(t)=b v(t) \,\rho_{mob}(t) ,
\end{eqnarray}
with the increment of plastic strain per time step $\Delta t$: $\Delta\epsilon_p(t) = \dot{\epsilon}_p(t) \Delta t.$
The corresponding applied shear stress $\tau_{app}$  is given by Hooke's law of elastic deformation,
\begin{equation}\label{eq:Hooke}
{\tau}_{app}(t) \equiv \mu \epsilon_{e}(t) = \mu \big[ {\epsilon}_{app}(t) - {\epsilon}_{p}(t) \big].
\end{equation}
This completes the set of thermodynamic and kinetic equations of our minimal dislocation glide model developed for solid $^4$He.

\subsection{Computation and model parameters}

The coupled system of equations are solved numerically with a simple time-step evolution method, because
it is sufficient to employ the Euler forward method to obtain the time evolution of the dynamics.
In this work, we assume initial conditions $\epsilon_{app}(0)=\epsilon_p(0)=0$ and $\rho_{mob}(0)=\rho_{stored}(0)=0.5\rho_{tot}(0)$. 
First we solve Eq.~(7) for the increment of applied shear stress $\Delta{\tau}_{app}(t)$ at each time step $\Delta t$ for an increment of applied strain
$\Delta \epsilon_{app}(t) = \dot{\epsilon}_{app} \Delta t$ with given dc strain rate $\dot{\epsilon}_{app}$, based on the difference between applied strain 
$\epsilon_{app}$ and plastic strain  $\epsilon_{p}$.  Then, we update the total applied shear stress $\tau_{app}(t)$ to get the average speed $v(t)$ of dislocations from Eqs.~(3) and (4). This way we solve the differential equation for the plastic strain rate $\dot\epsilon_{p}(t)$  in Eq.~(6), which implicitly depends on $\epsilon_{p}$, $\tau_{app}$ and $\tau_{int}$. Finally, we update the increment of the total dislocation density $\rho_{tot}(t)$ in Eq.~(2) and the interaction stress $\tau_{int}(t)$ in Eq.~(5).

In the calculations shown hereafter the pinning potential $U_i$ is set by the melting temperature $T_m= 1.86$ K with the assumption that the pinning is from dislocation crossing or local melting. This strongly reduced value takes into account the quantum fluctuations, because in classical crystals $U_i$ is of the order of the Debye temperature, which is roughly 26 K for hcp $^4$He.
For simplicity, we further assume that the Burgers vector is of the same magnitude as the lattice constant, $b=0.364$ nm, which would correspond to glide on either the basal or prismatic planes of the hcp  crystal. We approximate
the Debye frequency as $\nu_D=\Theta_D k_B/h=$ 600 GHz with a corresponding sound speed of 
$b\nu_D = 218$ m/s and shear modulus $\mu = 13.7$ MPa.\cite{Suzuki77}  
For the dislocation interaction coefficient, we use $\alpha$ = 0.01. 
This value is much smaller than for the hcp  metal Zr,
$0.1< \alpha < 2$,\cite{Beyerlein08,Lavrentev1980} or fcc Cu, $0.5 \alt \alpha \alt 1$ at room temperature.\cite{Mecking1981,Kocks03} 
A strongly reduced value should be expected when quantum fluctuations and interactions are important.
In addition, it takes into account dislocation screening effects as well as the screening due to $^3$He atoms bound to the dislocation line. 

The Peierls barrier is inferred from the recent ac shear displacement experiments in which the softening happens at an applied strain larger than 
$2.0 \times 10^{-8}$.\cite{Day07, Day09, Syshchenko10} 
By interpreting the softening of the solid as dislocation glide, we obtain $\tau_{Peierls} = 1.0 \times 10^{-8}$$\mu$. This value is consistent with the experiment by Sanders and coworkers,\cite{Sanders77}  who reported a nearly vanishing Peierls stress, whereas Hiki and Tsuruoka\cite{Hiki1983} reported a larger, but still small $\tau_{Peierls} \approx 10^{-5}$$\mu$. 
On the other hand, Monte Carlo simulations predicted much larger values for $\tau_{Peierls} \sim 0.006 - 0.03\mu$ for edge and screw dislocations, respectively.\cite{Pessoa2010}
Clearly the experimentally reported  values for the Peierls stress are unusually small, which might be the combined effects of quantum fluctuations, which play a role at low temperatures in reducing the barrier for the escape of immobile dislocations and the nucleation of double kinks or small loops of atomic size $b$ vs.\ $\Lambda$.\cite{Kocks75}
If in fact the activation volume for overcoming the Peierls barrier is on the order of $V=b^3$ rather than $V=b^2 \Lambda$, then the Peierls stress is effectively reduced by approximately $b/\Lambda$ in our formulation of the energy barrier $\Delta G$ in Eq.~(4).
Everything else the same, we can either write for the potential energy of the Peierls barrier $E_P = \tau_{Peierls} \times b^2\Lambda$ or $E_P = \tau_{Peierls}^{eff} \times b^3$. Consequently, $\tau_{Peierls}^{eff} \sim \tau_{Peierls} (\Lambda/b) = 10^{-8} \mu \, \Lambda/b \sim 0.03 \mu$ for typical dislocation density $\rho_{tot} = 1/\Lambda^2 = 10^{6}$ m$^{-2}$. This estimate of $\tau_{Peierls}^{eff}$ is also in reasonable agreement with the Monte Carlo simulations.\cite{Pessoa2010}
Thus a systematic study of the Peierls stress or critical strain versus dislocation density may resolve the {\it true} value of the Peierls stress.

Of course, all material parameters used above are subject to updates in experiments. We believe that the most realistic result can be obtained once the material parameters can be verified. In the next section, we will study how the plastic nature is revealed in the WHR and other experimental observables as well as a systematic study of the parameter dependence on the results.

\begin{figure*}[tb]
\rotatebox[origin=c]{0}{\includegraphics[width=0.75\textwidth]{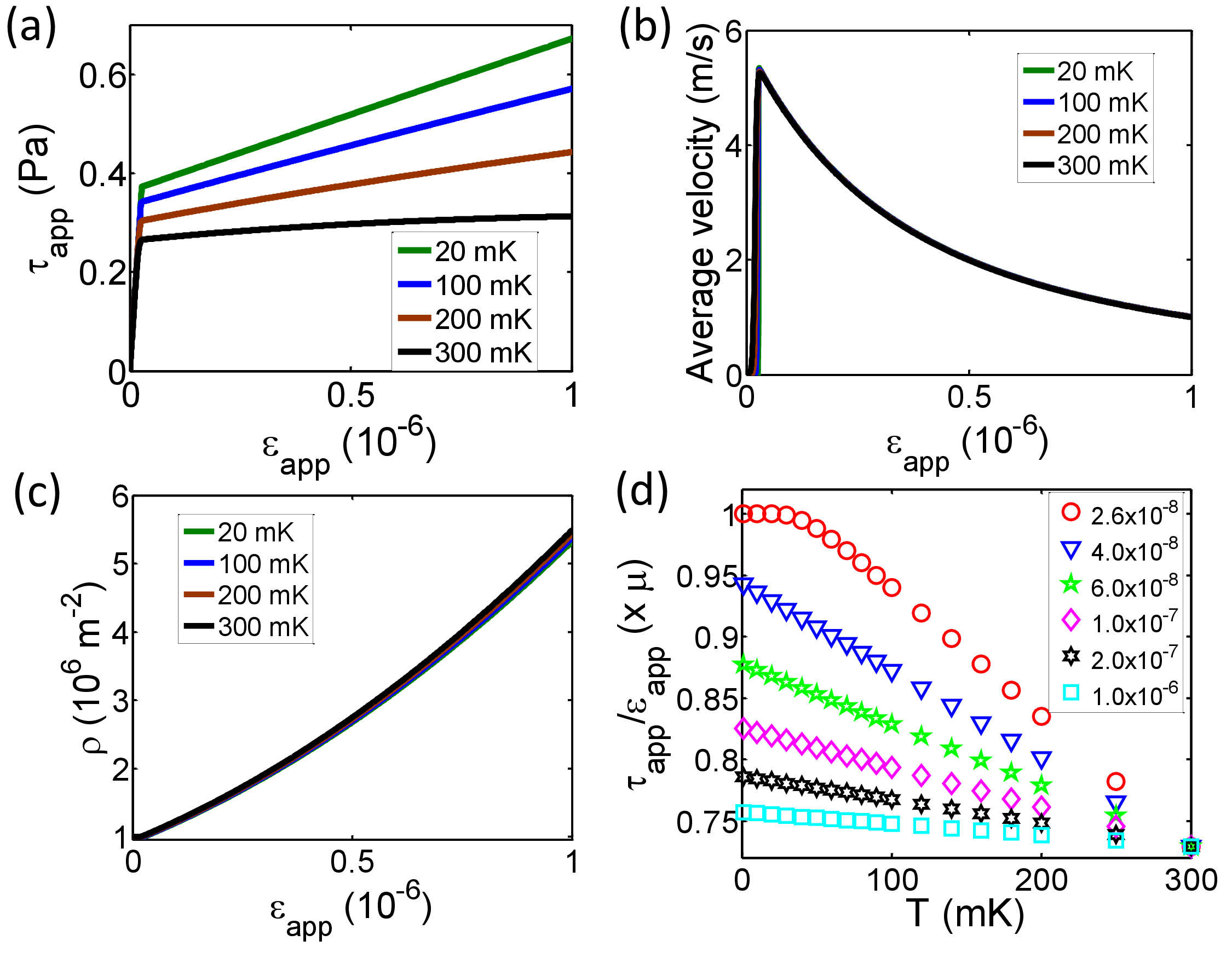}}
\caption{(color online).
(a) Stress-strain curves for different temperatures at an applied strain rate of $\dot\epsilon_{app}=10^{-3}$ s$^{-1}$ with initial total and mobile dislocation densities of $\rho_{tot}=1.0 \times 10^6$ m$^{-2}$ and $\rho_{mob}=0.5 \times 10^6$ m$^{-2}$. 
(b) Plot of the corresponding average speed-strain curves of (a) are nearly identical at different temperatures.
(c) Plot of the corresponding dislocation density-strain curves of (a) are nearly identical at different temperatures.
(d) The ``effective'' shear modulus $\mu_{eff}=\tau_{app}/\epsilon_{app}$ versus temperature; data for different strains have been shifted to coincide at 300 mK.
}
\label{fig:FIG2}
\end{figure*}

\section{Results and discussion }

\subsection{Strain-stress curves}

Dislocation glide manifests itself macroscopically as a change in the WHR. Prior to the onset of dislocation glide, the solid behaves elastically and the WHR equals the elastic shear modulus. At sufficiently large applied deformation, some of the pre-existing stored dislocations become unpinned and glide in reaction to the external stress. The glide of dislocations accommodates the deformation strain and the slope of the stress-strain response deviates from the shear modulus noticeably. The point of departure on the stress-strain response is normally referred to as the yield point, with corresponding yield strain and yield stress. Above the yield point is the so-called {\it plastic} regime. 
In this study, the yield point of the stress-strain curve is defined as the point where the ratio of applied stress over applied strain equals $0.99\mu$. Although this criterion is arbitrary, it is easy to implement and does not change any of the key results or conclusions. 

The calculated results are in qualitative agreement with recent ac shear experiments.\cite{Day07, Day09, Syshchenko10} 
In Fig.~\ref{fig:FIG2}(a) we plot the predicted shear stress-strain curves for solid $^4$He deformed under different temperatures with an initial dislocation density $\rho_0 = 1.0 \times 10^6$ m$^{-2}$. \cite{Rojas11} We find that yield strains at all temperatures are roughly the same value of $2.6 \times 10^{-8}$, which is close to the critical strain $2.2 \times 10^{-8}$ in the ac experiments. 

The temperature effect is most apparent in the stress-strain curve above the yield stress. At high temperature the WHR approaches zero, which means that the material undergoes great plastic deformation even with minimum applied stress. 
In Fig.~\ref{fig:FIG2}(b) and (c), the temperature effect on the evolution of the average speed and dislocation density are substantially weaker
than the ones shown for the parameters  in panel (a).
There are two reasons for this difference: 
(1) The dislocation density in panel (c) is at the level of $10^6$ m$^{-2}$; thus, a small difference on the order of $10^{6}$ m$^{-2}$ still has a strong influence on the plastic strain rate calculated from Eq.~(6), especially when the applied strain rate is $\dot\epsilon_{app}=10^{-3}\,{\rm s}^{-1}$. 
(2) The difference in stress-strain curves in Fig.~\ref{fig:FIG2}(a) is a consequence of the accumulated difference of plastic strain rates from each time step. Thus, the temperature effect is much more prominent in the stress-strain curves.

One can visualize the microscopic processes of the deformation through macroscopic quantities. The average dislocation speed $v$ is of the order of $\sim 2$ m/s or 1\% of the speed of sound and is plotted in Fig.~\ref{fig:FIG2}(b). It shows that below the yield point $v$ is zero and a boost is found right at the yield point. 
A slow decrease in $v$ above the yield strain occurs because the increasing dislocation density increases the mean-field interaction and sets a higher effective barrier for gliding.  Notice that although the ``average" dislocation speed decreases, there are still fast dislocations to accommodate the increasing external strain; this has been seen in an earlier study using discrete dislocation dynamics.\cite{Wang07} On the other hand, the increase in dislocation density $\rho$,
see Fig.~\ref{fig:FIG2}(c), is not as dramatic as for $v$, which we found to be true for all of our calculations.

As seen in Fig.~\ref{fig:FIG2}(a), the model also predicts a strong temperature dependence for the stress-strain curve, which follows directly from the kinetics embedded in the model of Eq.~(2).  The rate of dislocation motion across the solid is assumed to be controlled by thermally activated glide, Eq.~(2), and therefore, higher temperatures facilitate dislocation glide, enabling dislocations to glide faster at lower stresses. 

To summarize and allow direct comparison with experiments, we re-plot  Fig.~\ref{fig:FIG2}(a) as an effective shear modulus
$\mu_{eff}\equiv \tau_{app}/ \epsilon_{app}$ in Fig.~\ref{fig:FIG2}(d). Note that this is the definition used in the ac shear experiments, rather than the WHR, irrespective of the value of the applied strain.
One can see that the $\mu_{eff}$ curve decreases with increasing temperature, while the zero-temperature value $\mu_{eff}(0)$ decreases with increasing applied strain, which is in qualitative accordance with experiments. It is encouraging that our dc strain-rate model actually captures the essential features of the ac shear modulus experiments.

\begin{figure*}[htb]
\rotatebox[origin=c]{0}{\includegraphics[width=0.75\textwidth]{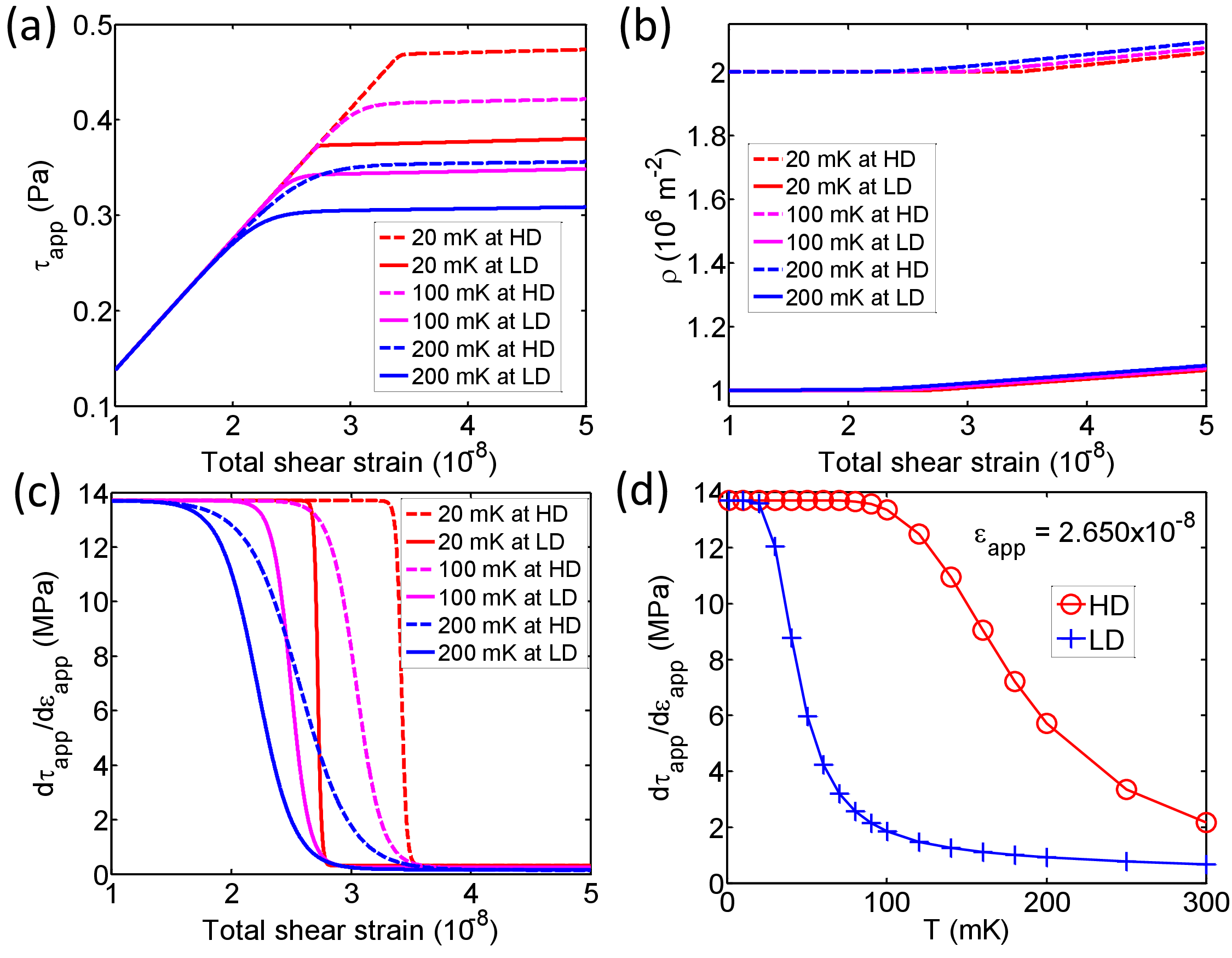}}
\caption{(color online).
(a) Stress-strain curves for high (HD) and low (LD) initial dislocation densities at $\dot\epsilon_{app}=10^{-3}$ s$^{-1}$.
(b) Plot of the corresponding dislocation density-strain curves of (a).
(c) Plot of the corresponding WHR-strain curves of (a).
(d) The WHR versus $T$ for HD and LD cases. The curves for HD and LD cases correspond to the initial total dislocation density of $1.0 \times 10^6$ m$^{-2}$ and $2.0 \times 10^6$ m$^{-2}$, respectively. All data are taken from the same strain amplitude at $\epsilon_{app}=2.650 \times 10^{-8}$.
}
\label{fig:FIG3}
\end{figure*}

\subsection{Dislocation density dependence}

\begin{figure}[htb]
\rotatebox[origin=c]{0}{\includegraphics[width=1.0\columnwidth]{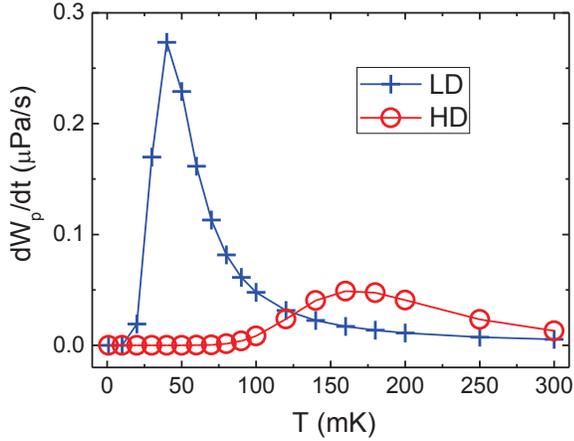}}
\caption{(color online).
Plastic dissipation curves corresponding to the WHR curves of Fig.~\ref{fig:FIG3}(d) for low (LD) and high (HD) densities of dislocations.
}
\label{fig:FIG4}
\end{figure}

We show through varying the initial dislocation density that our model describes reported annealing experiments.\cite{Day09}  Dislocations are created during the growth process and can either be increased or decreased by pre-stressing or annealing. In the calculation we focus on two initial dislocation line densities of high density (HD), $2.0 \times 10^6$ m$^{-2}$, and low density (LD), $1.0 \times 10^6$ m$^{-2}$, which correspond to an as-grown and annealed sample, respectively. We see again the plastic behavior above the yield point.  The yield point is higher for the HD case, see Fig.~\ref{fig:FIG3}(a). This can be understood since the larger mean-field interaction in the HD sample possesses a stronger resistance to gliding, because with increasing density the interaction stress $\tau_{int}$ increases as described by Eq.~(3). Notice that the increase in dislocation density after applying external strain is negligible, as shown in Fig.~\ref{fig:FIG3}(b).
This behavior is thus mainly determined by the initial dislocation density. In general the HD case also shows a larger crossover zone from the elastic to plastic regime, since a larger activation is required for dislocation lines to glide. We also compare the temperature effect on the two different density samples. In the temperature range of interest, 20-200 mK, the distinction between the two densities is clear. To better see the temperature effect, we plot the WHR as a function of temperature (Fig.~\ref{fig:FIG3}(d)) at a given applied strain, $\epsilon_{app}=2.65 \times 10^{-8}$. The WHR of the HD case deviates from the zero-temperature shear modulus at higher temperatures, approximately at 100 mK, compared to the LD scenario, which deviates at approximately 20 mK. This is because the larger dislocation interaction sets a higher barrier that requires a higher temperature to thermally activate glide. 
Again it is encouraging to see that our simple dislocation glide model captures the essential features of the ac shear experiments between as-grown and annealed samples.\cite{Day09}  

\begin{figure*}[tb]
\rotatebox[origin=c]{0}{\includegraphics[width=0.75\textwidth]{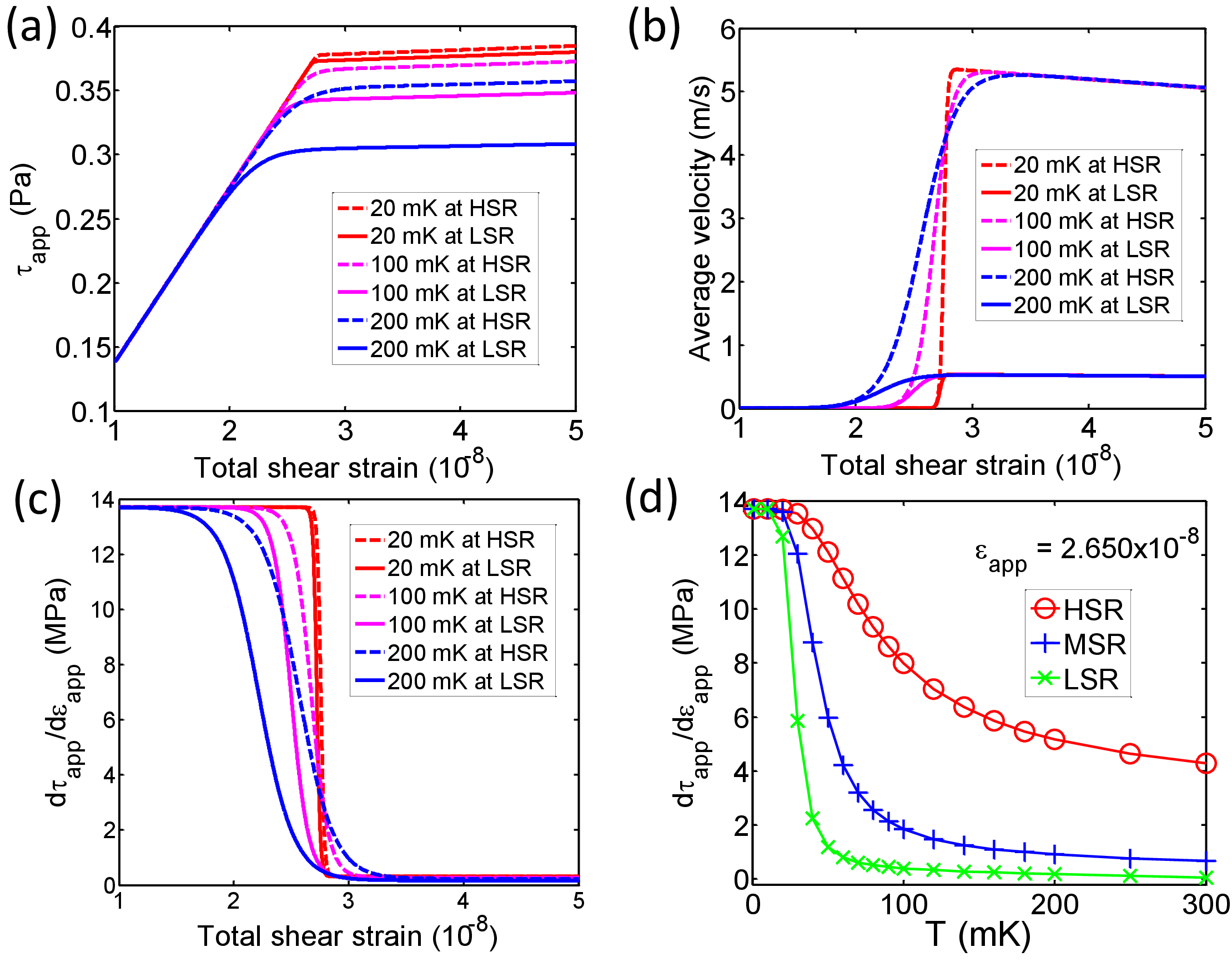}}
\caption{(color online).
(a) Stress-strain curves for different applied strain rates at $\dot\epsilon_{app}=10^{-3}$ s$^{-1}$;  
(b) plot of the corresponding average speed-strain curves of (a);   
(c) plot of the corresponding work-hardening rate-strain curves of (a); 
(d) work-hardening rate, WHR, versus temperature for different applied strain rates. (The curves for high and low applied strain rates correspond to $10^{-2}$ and $10^{-4}$ s$^{-1}$, respectively). All data are taken from the same strain amplitude at $\epsilon_{app}=2.650 \times 10^{-8}$.
}
\label{fig:FIG5}
\end{figure*}

\subsection{Dissipative glide}

Dissipation is one of the essential features of dislocation glide described in this paper. To overcome barriers and obstacles when gliding, a dislocation dissipates energy 
$W_p = \tau_{app} \,\Delta\epsilon_p$
per time step $\Delta t$. The corresponding dissipation rate is expressed as 
$\dot{W}_p$. 
From this expression it is easy to see that significant dissipation or damping happens only in the plastic regime above the yield point $\tau_y$,
where ${\epsilon}_p > 0$.  In general, a dissipation peak accompanies a drastic change in the WHR as a function of temperature. 
The plastic dissipation corresponding to the WHR in Fig.~\ref{fig:FIG3}(d) is shown in Fig.~\ref{fig:FIG4}. Here, the dissipation peak is centered near the temperature, where the WHR, 
$d\tau/d\epsilon$, is changing most rapidly, see Fig.~\ref{fig:FIG3}(d). 
This behavior is similar to the one seen in the inverse quality factor $1/Q$ of ac shear experiments.
In addition, the position of the peak  shifts to lower temperatures as the dislocation density decreases from the HD to LD scenario and the height of the peak increases with decreasing dislocation density. This narrowing of the dissipation peak can be understood  in terms of fewer dislocations in the system  resulting in fewer mobile dislocations with larger mean-free path to move. Again, that in turn causes a faster release of stored elastic energy in the system.

\subsection{Strain rate dependence}

Next, we study the effect of the key control parameter in experiment on the plastic response of solid $^4$He,
namely, the applied strain rate $\dot{\epsilon}_{app}$. 
Thermally activated rate-controlled dislocation glide implies a sensitivity to the applied strain rate. Two different strain rates were used in the following calculations:  a low strain rate (LSR) of $10^{-4}\,{\rm s}^{-1}$ and a high strain rate (HSR) of $10^{-2}\,{\rm s}^{-1}$.  
In Fig.~\ref{fig:FIG5}(a)-(c) we compare the model results for low and high strain rates. From Fig.~\ref{fig:FIG5}(a) we can see that increasing the strain rate increases the yield stress and yield strain. In other words, as the strain rate increases, the transition from the elastic to plastic regime is delayed.

This delay is due to the failure of plastic deformation to respond fast enough  (at each time step) to the large applied strain rate for HSR. The plastic strain rate, $\dot{\epsilon}_p \propto v$, is roughly ten times larger for HSR (Fig.~\ref{fig:FIG5} (b)) while the applied strain is in fact a hundred times larger. The plastic deformation, therefore, does not affect the applied stress much until a much larger strain is applied, see Eq.~(\ref{eq:Hooke}); the yield point then happens at a higher drive. 

Further, we observe that the sensitivity of the yield stress and strain to strain rate is enhanced at higher temperatures. Since the yield stress in the high strain rate case is higher, the driving force for dislocation glide is larger, causing the dislocations on average to speed up as shown in Fig.~\ref{fig:FIG5}(b).  
This sensitivity can also be seen in the variation in the WHR,
$d\tau/d\epsilon$, shown in Fig.~\ref{fig:FIG5}(c).  Here we see that the extent of the elastic regime is wider at the higher temperature.  Fig.~\ref{fig:FIG5}(d) compares the $d\tau/d\epsilon$ from high to low temperatures at different strain rates.  Two features are revealed.  First, changing the strain rate does not affect the low temperature behavior, i.e., below 20 mK.  Second, compared to the HSR case, the WHR decreases more rapidly in the LSR case, resulting in a larger drop of the WHR at the same amount of applied strain.  
Notably, the latter feature can also be found in the ac shear experiments, \cite{Day07, Syshchenko10} 
although the impact on the WHR is not as obvious as in our calculations. The difference may be related to the usage of $\mu_{eff}$ rather than the WHR in experiments or the different loading methods in ac versus dc experiments or the boundary conditions. \cite{Zhou12}

\subsection{Model parameter dependence}

Finally, it is important to characterize the material parameter dependence of our model results. A suitable experimentally observable quantity for that purpose is the yield point. We study systematically the effect of the pinning potential $U_i$, the Peierls barrier $\tau_{Peierls}$, and the dislocation interaction strength $\alpha$ at 20 mK and 300 mK, since they play important roles in determining the average dislocation speed $v$ given in Eq.~(3) and the WHR. We keep all model parameters the same as in Sec.~III.A, except for the parameter whose dependence on the yield stress is investigated.

\begin{figure}[htb]
\rotatebox[origin=c]{0}{\includegraphics[width=1.0\columnwidth]{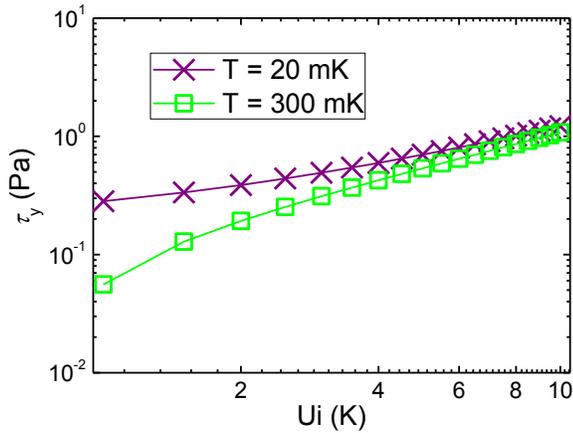}}
\caption{(color online).
Plot of yield stress $\tau_y$ versus pinning potential $U_i$ at both low (20 mK) and high (300 mK) temperatures.
}
\label{fig:FIG6}
\end{figure}

\begin{figure}[htb]
\rotatebox[origin=c]{0}{\includegraphics[width=1.0\columnwidth]{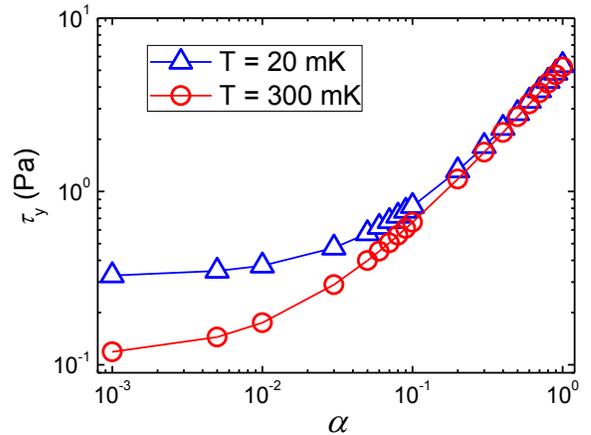}}
\caption{(color online).
Plot of yield stress $\tau_y$ versus dislocation interaction strength $\alpha$ at both low (20 mK) and high (300 mK) temperatures.
}
\label{fig:FIG7}
\end{figure}

\begin{figure}[htb]
\rotatebox[origin=c]{0}{\includegraphics[width=1.0\columnwidth]{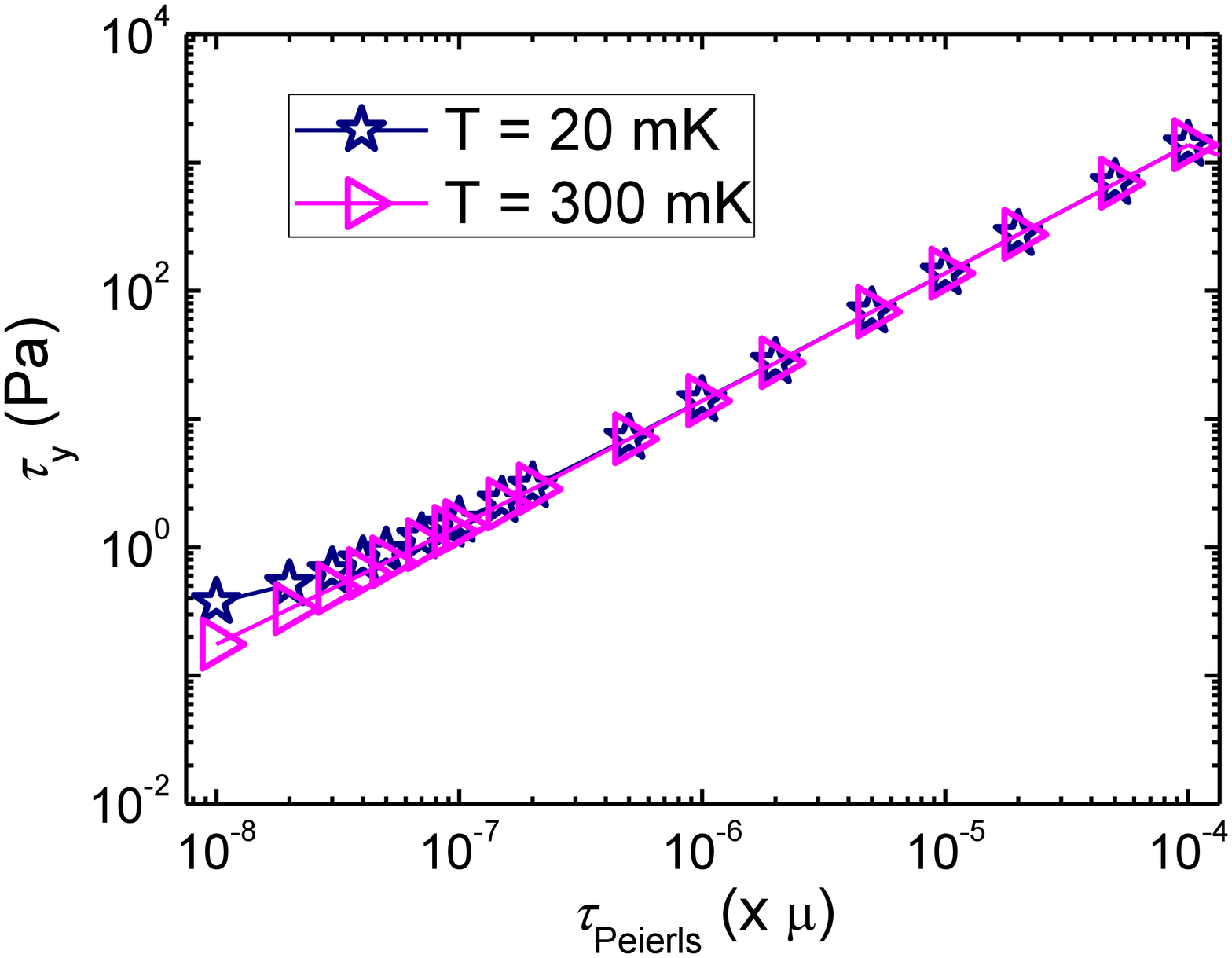}}
\caption{(color online).
Plot of yield stress $\tau_y$ versus pinning potential $\tau_{Peierls}$ at both low (20 mK) and high (300 mK) temperatures.}
\label{fig:FIG8}
\end{figure}

All parameters represent physical barriers to dislocation glide. Accordingly, the yield stress $\tau_y$ increases with all three parameters as expected (Fig.~\ref{fig:FIG6}, \ref{fig:FIG7}, \ref{fig:FIG8}). Among the three parameters, the dependence  on $U_i$ is least crucial, see Fig.~\ref{fig:FIG6}, since the value of $U_i$ is known to be restricted by the melting temperature $T_m=1.86$ K.
(1) The yield stress $\tau_y$ varies by less than an order of magnitude in the given range of $U_i$ values with the scaling relation $\tau_y \sim 0.1\, U_i \, {\rm Pa \over K}$. 
(2) The interaction strength $\alpha$, although difficult to control in experiments, is believed to range between 0.1 to 2 when compared with reported values for ordinary metals. In our model calculations we chose $\alpha=0.01$ to properly describe the WHR. This strongly reduced value, compared to ordinary metals, might be caused by interdislocation screening effects originating from quantum fluctuations.
Our calculation shows that $\tau_y$ varies by roughly an order of magnitude for $\alpha$ between 0.01 and 1, 
see Fig.~\ref{fig:FIG7}. The curve saturates below $\alpha<0.01$, because then $\tau_y$ is dominated by other barrier types, e.g., Peierls barrier. For $\alpha > 10^{-2}$ an approximate scaling of $\tau_y \approx (0.14 + 5  \alpha) \,{\rm Pa}$ can be inferred.
(3) For the Peierls potential, however, the yield stress changes by more than three orders of magnitude over the reported range shown in Fig.~\ref{fig:FIG8}. It satisfies a simple scaling relation $\tau_y \propto 1.4\cdot 10^7 \,  \tau_{Peierls}/\mu\, {\rm Pa} \approx \tau_{Peierls}$, which can be used to estimate the order of magnitude of the Peierls barrier.
From this work it has become clear that a reliable and independent determination of the Peierls barrier is needed. On the other hand, increasing temperature accentuates the parameter dependence on $\tau_y$ in general, because the solid is softer at higher temperatures. Fortunately, the temperature dependence of $\tau_y$ versus $\tau_{Peierls}$, the least well-known parameter in our dislocation model,  is negligible.          

\section{Conclusion}

In summary, we proposed and studied a model of dc strain rate driven dislocation glide in solid $^4$He.  We predict stress-strain curves and work-hardening rates as well as the role of the dislocation interactions and Peierls barriers. 
We find that the effective shear modulus (ratio of stress over strain) can exhibit, what appears to be, anomalous elasticity. 
Based on our predictions for the effective shear modulus and work-hardening rate, this anomaly is caused by the free glide of dislocations. 
It can be described rather well by classical theories of plastic deformation with renormalized model parameters due to zero-point fluctuations of the quantum solid. 
No supersolidity or quantum mechanism is required to explain these effects.

One may speculate about the realm of quantum and super plasticity in the quantum solid helium. So far the results of our 
classical thermally activated dislocation glide model would suggest that quantum tunneling of dislocation lines has not yet been seen in experiments. 
Indeed, measurements of the dynamic shear modulus down to 20 mK confirm the dominance of thermally activated relaxation behavior in the dissipation process.\cite{Syshchenko10,Su2010,Su2011}

Our predictions for the strain-stress curve in dc strain rate measurements are obtained with very reasonable parameter values. Since the parameter range is generally constrained, the model gives rather robust predictions for the stress-strain curve, which is encouraging particularly in the event that some of the parameters 
need to be updated with the advent of more accurate materials characterization. Future dc experiments can certainly narrow down the range of values for some of the model parameters. For example, our model calculations show over a wide range simple linear relations between yield stress and Peierls stress, $\tau_y \propto \tau_{Peierls}$, yield stress and pinning potential, 
$\tau_y \propto U_i$, and yield stress and screening strength of the dislocation interaction, $\tau_y \propto \alpha$. 
In addition, we find a direct relationship between the work-hardening rate and the applied strain rate as well as the dislocation density that resembles the effects of frequency and annealing process reported in ac shear strain experiments. Furthermore, using current parameters, our stress-strain curve predictions reveal that dislocations glide freely with an average speed of 1 to 5 m/s, which is about 1\% of the speed of sound in solid $^4$He and significantly faster than estimates based on the Granato-L\"ucke theory of vibrating dislocations would suggest.\cite{Haziot2013b}

Although the current dislocation glide model does not predict hysteresis, it can be extended, in principle, to capture the irreversible processes of creation and annihilation of dislocations during loading and reverse loading. 
We encourage and welcome experiments that will allow a more precise dynamical and structural characterization of the quantum solid $^4$He to provide stringent tests on quantum vs.\ classical dislocation glide at low temperatures.

\acknowledgments

We acknowledge helpful discussions with C.~J.\ Olson-Reichhardt, J. Toner, and Z. Nussinov.
This work was supported by the
U.S.\ DOE at Los Alamos National Laboratory under contract No.~DE-ac52-06NA25396 through the Office of Basic Energy Sciences, Division of Materials Sciences and Engineering.
C.~Z.\ acknowledges support by the Center for Nonlinear Studies through the LANL LDRD program
and the Materials Research Center at Missouri S\&T.

\section*{Supplemental Material}

The animation accompanying the Supplemental Material and depicted in the Fig.~(1) of the main text shows the results from a three-dimensional discrete dislocation dynamics (3D DDD) simulation.\cite{s1,s2} 
In the 3D DDD simulation, the dislocation lines have been discretized into segments, in which the dislocations are the simulated entities for studying the effects of plasticity.
Recently, this method has been used to examine the role of dislocation multiplication and mobility on the plasticity in small samples. Application was made to single-crystal fcc Ni to compute the elastic field, forces, and motion of dislocation loops.\cite{s2}

\end{document}